\documentclass[aps,prc,reprint,groupedaddress]{revtex4-2}

\usepackage[T1]{fontenc}
\usepackage[utf8]{inputenc}

\usepackage{mathptmx}
\usepackage{amsmath}
\usepackage{amssymb}
\usepackage{upgreek}
\usepackage{moresize}
\usepackage{microtype}
\usepackage{graphicx}

\begin{document}

\title{Spatial resolution of dijet photoproduction in near-encounter
ultraperipheral nuclear collisions}

\author{Kari J.\ Eskola}
\author{Vadim Guzey}
\author{Ilkka Helenius}
\author{Petja Paakkinen}
\email{petja.k.m.paakkinen@jyu.fi}
\author{Hannu Paukkunen}
\affiliation{University of Jyväskylä, Department of Physics, P.O. Box 35, FI-40014 University of Jyväskylä, Finland\\ and Helsinki Institute of Physics, P.O. Box 64, FI-00014 University of Helsinki, Finland}

\date{August 19, 2024}

\begin{abstract}
  We present next-to-leading order perturbative QCD predictions for inclusive dijet photoproduction in ultra-peripheral nucleus-nucleus collisions (UPCs) within the impact-parameter dependent equivalent photon approximation. Taking into account the finite size of both the photon-emitting and the target nucleus, we show that this process is sensitive to the transverse-plane geometry of the UPC events. We show that this leads to a sizeable, 20-40\% effect for large values of the $z_\gamma$ variable in the dijet photoproduction cross section in lead-lead UPCs at 5.02 TeV compared to the widely-used pointlike approximation where the nuclear radius is accounted for only as a sharp cut-off in the photon flux calculation. This resolution of the spatial degrees of freedom is a result of having high-transverse-momentum jets in the final state, which at the large-$z_\gamma$ kinematics requires a highly energetic photon in the initial state, thus biasing the collisions to small impact-parameter ``near-encounter'' configurations. We further discuss the role of the forward-neutron event-class selection in isolating the photonuclear cross section in the nucleus-nucleus collisions, and employ the needed electromagnetic breakup survival factor in our predictions.
\end{abstract}

\maketitle

\section{Introduction}

In ultra-peripheral nucleus-nucleus collisions (UPCs), two nuclei pass each other at an impact parameter large enough to prevent hadronic interaction via the strong force, but close enough so that they can interact though their electromagnetic (e.m.) fields~\cite{Bertulani:2005ru,Baltz:2007kq}. For collisions at high enough energy, these e.m.\ fields are strongly Lorentz-contracted and can be viewed in a so-called equivalent photon approximation (EPA)~\cite{Fermi:1924tc,vonWeizsacker:1934nji,Williams:1934ad} as a flux of quasi-real photons moving along the beam direction. This enables one to study photonuclear processes, where it becomes possible to produce e.g.\ high-transverse-momentum jets through photon-parton scattering, providing a probe on the partonic structure of nuclei~\cite{Strikman:2005yv,Guzey:2018dlm,Guzey:2019kik}. A primary interest in measuring this process inclusively~\cite{ATLAS:2017kwa,ATLAS:2022cbd} has therefore been to provide new constraints for the nuclear parton distribution functions (nPDFs), reviewed recently in Ref.~\cite{Klasen:2023uqj}. By requiring that both of the involved nuclei stayed intact in the process, it is also possible to study the previously unconstrained diffractive nPDFs and factorization-breaking effects in hard diffraction~\cite{Guzey:2016tek,Helenius:2019gbd}.

Previous studies at next-to-leading order (NLO) in perturbative QCD (pQCD) on the impact of UPC dijets on nPDFs~\cite{Guzey:2018dlm,Guzey:2019kik} have resorted to the very convenient pointlike approximation of the (effective) photon flux. For photons of very high energy, this might however not be accurate due to geometrical effects originating from these high-energy photons being emitted predominantly at small impact parameters~\cite{Baron:1993nk}. In this article, we will show that these geometrical effects become sizeable in the hard inclusive UPC dijet production, where the presence of high-transverse-momentum jets in the final state requires an energetic photon in the initial state. In particular, we demonstrate that one has to take into account the finite size of the target nucleus in a way that is not realised with a simple cut in the impact parameter of the two crossing nuclei. However, as we find out, this spatial resolution is not strong enough to probe the spatial dependence of the nuclear modifications in nPDFs.

As a further improvement over the previous NLO predictions, we take into account here the selection of forward-neutron event class, which is needed experimentally to isolate the photonuclear dijet cross section. For this, we include in our prediction the probability of no e.m.\ breakup of the photon-emitting nucleus~\cite{Baltz:2002pp}. We show that the geometrical effects mentioned above survive even after including this additional (and substantial) suppression factor. We discuss also the relation of this neutron-class selection to the diffractive contribution in the inclusive dijet cross section.

The rest of this article is organised as follows. In Section~\ref{sec:impact-par-dep-factorization}, we discuss how the geometrical effects arise in the photonuclear UPC processes and define the related impact-parameter dependent factorization of the UPC dijet cross section. In Section~\ref{sec:eff-flux}, we then define the effective, geometrically modified, photon flux and discuss different approximations for it. With these tools at hand, we provide in Section~\ref{sec:ATLAS-UPC-dijets} predictions in NLO pQCD for the dijets in lead-lead (Pb+Pb) UPCs at 5.02 TeV within the kinematics of the ATLAS preliminary results~\cite{ATLAS:2022cbd}. Subsequently, in Section~\ref{sec:bu-class}, we discuss the e.m.\ breakup modelling necessitated by the experimental neutron-class selection. Finally, in Section~\ref{sec:spatial-npdfs}, we discuss the resolving power of this observable in terms of the spatially dependent nuclear modifications, and provide a summary of the findings in Section~\ref{sec:summary}.

\section{Impact-parameter dependent factorization}
\label{sec:impact-par-dep-factorization}

\begin{figure}
  \includegraphics{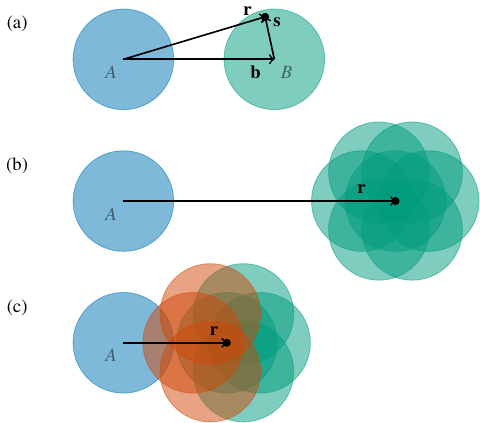}
  \caption{(a)~Specification of the transverse-plane vectors: $|{\bf b}|$ is the impact parameter between the nuclei $A$ (blue) and $B$ (green), $|{\bf r}|$ is the distance from the centre of the emitting nucleus $A$ to the point at which the emitted photon interacts with the nucleus $B$, and $|{\bf s}| = |{\bf r} - {\bf b}|$ is the distance from the centre of the nucleus $B$ to that same point. (b)~Different impact-parameter configurations for a fixed ${\bf r}$. When the distance $|{\bf r}|$ is larger than $R_A + 2R_B$, the impact-parameter vector ${\bf b}$ can assume any configuration that leads to a photonuclear interaction ($|{\bf s}| < R_B$) without causing hadronic interaction to occur. (c)~As the radius $|{\bf r}|$ gets smaller, some impact-parameter configurations ($|{\bf b}| < R_A + R_B$, marked in red color) lead to a nuclear overlap, and are thus excluded in UPCs.}
  \label{fig:transverse_plane}
\end{figure}

A transverse-plane view of a photonuclear UPC process is shown in FIG.~\ref{fig:transverse_plane}~(a). We identify $A$ as the photon-emitting and $B$ as the target nucleus. We assume the nuclei to be spherically symmetric such that their orientation does not play a role here. These nuclei pass each other at an impact parameter $|{\bf b}|$ larger than the sum of their radii $R_A + R_B$, thus suppressing any hadronic interaction. The interaction is then mediated in EPA by a quasi-real photon, whose flux is evaluated at a transverse distance $|{\bf r}|$ from the center of the nucleus $A$. This photon then probes the nucleus $B$ at a distance $|{\bf s}| = |{\bf r} - {\bf b}|$ from its center.

For a typical UPC process producing a low-mass final state with limited ``hardness'', the cross section gets a dominant contribution from large-impact-parameter configurations where $|{\bf r}| \sim |{\bf b}| \gg |{\bf s}|$. Such a ``far-passing'' scenario is decipicted in FIG.~\ref{fig:transverse_plane}~(b) for a fixed ${\bf r}$. In this region, any configuration of ${\bf b}$ that leads to a photonuclear interaction ($|{\bf s}| < R_B$) is allowed and thus the ${\bf s}$-dependence effectively integrates out. Conversely, if $|{\bf r}| \sim |{\bf b}| \sim |{\bf s}|$, some events will pass the UPC requirement $|{\bf b}| > R_A + R_B$, but there are also configurations which lead to nuclear overlap, as shown in FIG.~\ref{fig:transverse_plane}~(c), and are thus excluded from the UPC cross section. In this ``near-encounter'' region the full transverse-plane geometry can thus become important.

These geometrical effects were first discussed in Refs.~\cite{Baron:1993nk,Greiner:1994db,Krauss:1997vr}, where the impact-parameter dependent EPA is also derived. Here, we apply this formalism to the UPC dijet photoproduction, where the inclusive ($X, X'$ denote ``anything'') differential cross section can be written as
\begin{multline}
  {\rm d} \sigma^{AB \rightarrow A + {\rm dijet} + X} \\
  \begin{split}
    = \sum_{i,j,X'} & \int {\rm d}^2{\bf b} \, \Gamma_{AB}({\bf b}) \int {\rm d}^2{\bf r} \, f_{\gamma/A}(y,{\bf r}) \otimes f_{i/\gamma}(x_\gamma,Q^2) \\
    \otimes & \int {\rm d}^2{\bf s} \, f_{j/B}(x,Q^2,{\bf s}) \times \delta({\bf r}\!-\!{\bf s}\!-\!{\bf b}) \\
    \otimes &\ {\rm d} \hat{\sigma}^{ij \rightarrow {\rm dijet} + X'}(x_\gamma y p_A,x p_B,Q^2).
  \end{split}
  \label{eq:xsec_full}
\end{multline}
Here, $\Gamma_{AB}({\bf b})$ is the probability for \emph{no} hadronic interaction between the nuclei $A$ and $B$ at the impact parameter $|{\bf b}|$~\footnote{In the eikonal approximation, $\Gamma_{AB}({\bf b})$ can be expressed through the optical theorem in terms of the elastic nucleus $A$ - nucleus $B$ scattering amplitude.}. Depending on the experimental event selection, this factor should include also the probability for e.m.\ breakup, as we will discuss in detail in Section~\ref{sec:bu-class}. The (bare) photon flux $f_{\gamma/A}(y,{\bf r})$ is to be understood as the number of (quasi-real) photons, emitted from the nucleus $A$, per unit area at position ${\bf r}$. The function $f_{i/\gamma}(x_\gamma,Q^2)$ accounts here for both the photon parton distribution in the resolved case with $i = q,\bar{q},g$ and for the direct contribution with $i = \gamma$, where $f_{\gamma/\gamma}(x_\gamma,Q^2) = \delta(1 - x_\gamma)$. Finally, the impact-parameter dependent nuclear PDF $f_{j/B}(x,Q^2,{\bf s})$ encodes the transverse spatial distribution of partons $j$ in the nucleus $B$ as a function of the momentum fraction $x$ and the transverse distance $|{\bf s}|$ from the center of the nucleus, and ${\rm d} \hat{\sigma}^{ij \rightarrow {\rm dijet} + X'}$ is the partonic differential cross section for producing the dijet final state. The multiplicative convolutions $\otimes$ are understood to be over the momentum fractions $y$, $x_\gamma$ and $x$ such that the photon carries a momentum $k = y p_A$ and the partons $i$, $j$ have the momenta $x_\gamma y p_A$, $x p_B$, respectively, where $p_{A,B}$ are the per-nucleon beam momenta so that $s_{\rm NN} = (p_A + p_B)^2$ is the nucleon-nucleon center-of-mass (c.m.s.) energy. We have fixed here the factorization and renormalization scales to a single hard scale $Q$ associated with the jet transverse momenta. This scale is taken to be much larger than the incident photon or the intrinsic parton transverse momenta, which are therefore neglected.

Equation~\eqref{eq:xsec_full} relies on the space-time separation (factorization) of the processes of the electromagnetic emission of equivalent photons, accompanied by the restriction on soft nucleus-nucleus interaction, from the dijet production in the strong photon-nucleus interaction. The latter is treated in the framework of the collinear factorization of perturbative QCD at the next-to-leading order accuracy, where the dijet cross section is given by convolution of photon and nuclear PDFs (including the direct photon contribution) and the hard partonic cross section~\cite{RevModPhys.74.1221}. It is assumed that both types of factorization hold differentially in ${\bf b}$, i.e., they are applicable to the distributions depending on the impact parameter, as is necessary for applying the UPC condition through $\Gamma_{AB}({\bf b})$. Note that the Fourier transform of $f_{j/B}(x,Q^2,{\bf s})$ with respect to ${\bf s}$ relates it to the nuclear generalized parton distributions (GPDs) in the non-skewed limit~\cite{Burkardt:2000za,*Burkardt:2002hr}.

\section{Effective photon flux}
\label{sec:eff-flux}

In order to demonstrate the importance of the geometrical effects, we will start by assuming that one can factorize the impact-parameter dependent nPDFs as
\begin{equation}
  f_{j/B}(x,Q^2,{\bf s}) = \frac{1}{B} \, T_{B}({\bf s}) \times f_{j/B}(x,Q^2)
  \label{eq:b_dep_fact_assump}
\end{equation}
where $f_{j/B}(x,Q^2)$ is the ordinary spatially-independent (averaged) nPDF and $T_{B}({\bf s})$ is the nuclear thickness function describing the (parton) spatial distribution. We will revisit the validity of this assumption in Section~\ref{sec:spatial-npdfs}. With it, we can re-write Eq.~\eqref{eq:xsec_full} as
\begin{multline}
  {\rm d} \sigma^{AB \rightarrow A + {\rm dijet} + X} = \sum_{i,j,X'} f_{\gamma/A}^{\rm eff}(y) \otimes f_{i/\gamma}(x_\gamma,Q^2) \otimes f_{j/B}(x,Q^2) \\ \otimes {\rm d} \hat{\sigma}^{ij \rightarrow {\rm dijet} + X'}(x_\gamma y p_A,x p_B,Q^2)
  \label{eq:xsec_w_eff_flux}
\end{multline}
where
\begin{equation}
  f_{\gamma/A}^{\rm eff}(y) = \frac{1}{B} \int {\rm d}^2{\bf r} \int {\rm d}^2{\bf s} \, f_{\gamma/A}(y,{\bf r}) \, T_{B}({\bf s}) \, \Gamma_{AB}({\bf r}\!-\!{\bf s})
  \label{eq:eff_flux}
\end{equation}
is the \emph{effective} photon flux. This leads to a tremendous simplification in numerical calculations, where Eq.~\eqref{eq:xsec_w_eff_flux} can now be implemented directly in the dijet photoproduction codes and all the geometrical integrals can be precomputed in evaluating Eq.~\eqref{eq:eff_flux}. For a technical implementation, see the Appendix. These formulae coincide with the ones in Ref.~\cite{ATLAS:2022cbd}, where the effective flux was used to reweight Monte-Carlo predictions from Pythia~8. Note that while we denote the effective flux with a subscript ``${\gamma/A}$'' to indicate that the photon was emitted from nucleus $A$, it still carries information also on the target nucleus through the $T_{B}$ and $\Gamma_{AB}$ functions.

\begin{figure*}
  \centering
  \includegraphics{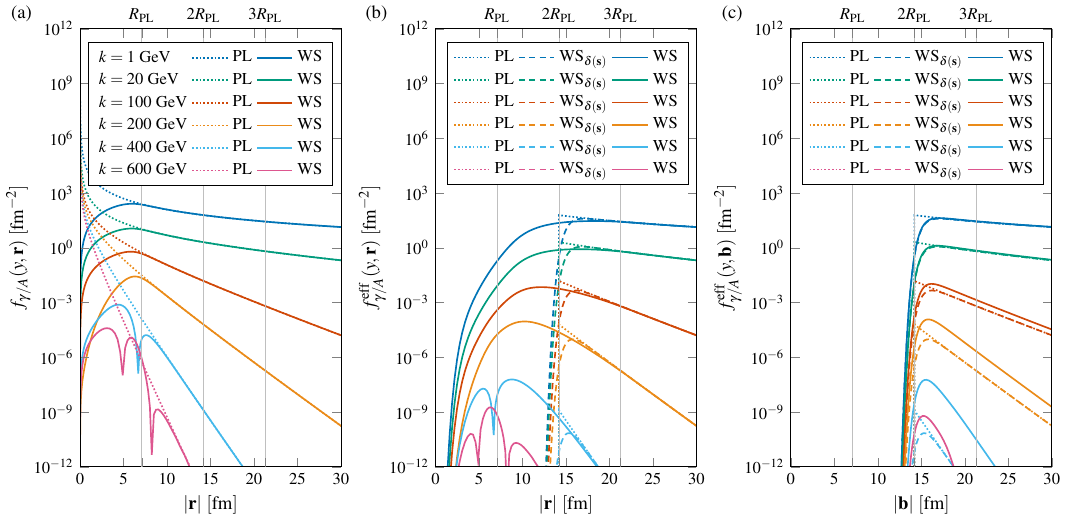}
  \caption{The bare and effective photon flux for different photon energies $k = y\sqrt{s_{\rm NN}}/2$ in the nucleon-nucleon c.m.s.~frame for Pb+Pb collisions at $\sqrt{s_{\rm NN}} = 5.02\ {\rm TeV}$. (a) A comparison of the bare photon flux from the PL~(Eq.~\eqref{eq:pl-bare-flux}) and WS~(Eq.~\eqref{eq:ws-bare-flux}) source. (b) A comparison of the $|{\bf r}|$-dependent effective flux, Eq.~\eqref{eq:eff_flux_r_dep}, from the three different approximations. (c) As previous, but now as a function of $|{\bf b}|$, cf.\ Eq.~\eqref{eq:eff_flux_b_dep}.}
  \label{fig:ph-flux}
\end{figure*}

We will now discuss this effective flux under different approximations, also in terms of more differential forms
\begin{align}
  f_{\gamma/A}^{\rm eff}(y, {\bf r}) &= f_{\gamma/A}(y,{\bf r}) \times \frac{1}{B} \int {\rm d}^2{\bf s} \, T_{B}({\bf s}) \, \Gamma_{AB}({\bf r}\!-\!{\bf s})
  \label{eq:eff_flux_r_dep} \\
  \intertext{and}
  f_{\gamma/A}^{\rm eff}(y, {\bf b}) &= \Gamma_{AB}({\bf b}) \times \frac{1}{B} \int {\rm d}^2{\bf s} \, f_{\gamma/A}(y,{\bf b}\!+\!{\bf s}) \, T_{B}({\bf s})
  \label{eq:eff_flux_b_dep}
\end{align}
which by definition integrate to Eq.~\eqref{eq:eff_flux},
\begin{equation}
  f_{\gamma/A}^{\rm eff}(y) = \int {\rm d}^2{\bf r} \, f_{\gamma/A}^{\rm eff}(y, {\bf r}) = \int {\rm d}^2{\bf b} \, f_{\gamma/A}^{\rm eff}(y, {\bf b}),
\end{equation}
but, as we demonstrate in Section~\ref{subsec:full_ws}, encode the spatial dependence in rather different ways.

\subsection{Point-like approximation}

In the simplest approximation, we take both the source and target nucleus as pointlike objects and exclude the events with nuclear overlap using a simple cut in the impact parameter at $b_{\rm min} \approx R_A + R_B$. In this pointlike (PL) approximation, the bare photon flux of a nucleus with charge $Z$ reads~\cite{Bertulani:1987tz}
\begin{equation}
  f_{\gamma/A}^{\rm PL}(y,{\bf r}) = \frac{Z^2 \alpha_{\rm e.m.}}{\pi^2} m_p^2 y [ K_1^2(\xi) + \frac{1}{\gamma_L} K_0^2(\xi) ], \quad \xi = y m_p |{\bf r}|,
  \label{eq:pl-bare-flux}
\end{equation}
where $K_{0,1}$ are modified Bessel functions of the second kind, $\gamma_L = \sqrt{s_{\rm NN}}/2m_p$ is the Lorentz factor and $m_p = m_A/A$ the mass of a single nucleon. We also have
\begin{equation}
  T_B^{\rm PL}({\bf s}) = B\delta^{(2)}({\bf s}), \quad \Gamma_{AB}^{\rm PL}({\bf b}) = \theta(|{\bf b}| - b_{\rm min}).
  \label{eq:pl-cond}
\end{equation}
For Pb+Pb collisions, we take $b_{\rm min} = 2R_{\rm PL} = 14.2\ {\rm fm}$ as in Ref.~\cite{Guzey:2018dlm}, where $R_{\rm PL} = 7.1\ {\rm fm}$ defines the nuclear hard-sphere radius. Integrating over the transverse-plane variables, the delta function in Eq.~\eqref{eq:pl-cond} forces the vectors ${\bf r}$ and ${\bf b}$ to be the same and one recovers the well-known analytic expression~\cite{Bertulani:1987tz}
\begin{multline}
  f_{\gamma/A}^{\rm eff,PL}(y) = \frac{2 Z^2 \alpha_{\rm e.m.}}{\pi y} \!\left[ \zeta K_0(\zeta) K_1(\zeta) \!-\! {\textstyle \frac{\zeta^2}{2}} [ K_1^2(\zeta) \!-\! K_0^2(\zeta) ] \right]\!\!, \\ \zeta = y m_p b_{\rm min},
\end{multline}
coinciding with the one used in Ref.~\cite{Guzey:2018dlm} for the UPC dijet photoproduction. We note that $b_{\rm min}$ is an effective parameter and could depend on the collision energy due to the increase of the nucleon-nucleon interaction radius at high energy.

The resulting photon fluxes (bare and effective) are presented as a function of $|{\bf r}|$ in FIG.~\ref{fig:ph-flux} (a and b) as dotted lines. The bare pointlike flux has a divergence at $|{\bf r}| = 0$, but this is excluded from the effective flux by the cut $|{\bf r}| > 2R_{\rm PL}$. We note that for large photon energies the flux drops extremely fast as a function of $|{\bf r}|$, even by few orders of magnitude over the diameter of a lead nucleus. Moreover, this behaviour is not linear, but rather almost exponential, which will have consequences later on.

\subsection{Improving the approximation}

The PL approximation given above is very useful for applications where the dominant contribution to the UPC cross section comes from large $|{\bf r}|$ but, as we will show, it is too crude to describe accurately phenomena at small $|{\bf r}|$. We will now improve this approximation step by step.

First, we can trade the pointlike approximation of the bare flux with the one for an extended charge distribution~\cite{Krauss:1997vr},
\begin{multline}
  f_{\gamma/A}^{\rm WS}(y,{\bf r}) = \frac{Z^2 \alpha_{\rm e.m.}}{\pi^2} \frac{1}{y} \times \\ \left| \int_0^\infty \frac{{\rm d}k_\perp k_\perp^2}{k_\perp^2 + (y m_p)^2} F_A^{\rm WS}(k_\perp^2 + (y m_p)^2) J_1(|{\bf r}|k_\perp) \right|^2,
  \label{eq:ws-bare-flux}
\end{multline}
where $J_1$ is the cylindrical modified Bessel function of the first kind and the form factor is taken as the Fourier transform
\begin{equation}
  \begin{split}
    F_A^{\rm WS}(t) &= \frac{1}{A} \int {\rm d}^3 \vec{r} \, \left.\rho_A^{\rm WS}(|\vec{r}|) \, {\rm e}^{{\rm i} \vec{q}\cdot\vec{r}} \, \right|_{|\vec{q}|=\sqrt{|t|}} \\
    &= \frac{4\pi}{A} \int_0^\infty {\rm d}r \left. \frac{r}{q} \sin(qr) \rho_A^{\rm WS}(r) \right|_{q=\sqrt{|t|}},
  \end{split}
\end{equation}
using the values $R_{\rm WS} = 6.49$ fm and $d_{\rm WS} = 0.54$ fm for the lead nucleus in the two-parameter Woods-Saxon (WS) distribution
\begin{equation}
  \rho_A^{\rm WS}(r) = \frac{\rho_0}{1 + \exp[(r-R_{\rm WS})/d_{\rm WS}]}
  \label{eq:ws-distr}
\end{equation}
normalised such that $F_A^{\rm WS}(0) = 1$. The bare flux obtained this way is shown in FIG.~\ref{fig:ph-flux}~(a) with solid lines. As is well known, the differences to the PL approximation reside almost exclusively in the region within the emitting nucleus. Only for the very high energy photons where additional diffractive-like minima appear due to the integration over the form factor we see differences at distances up to $|{\bf r}| \approx 10\ {\rm fm}$. With a cut at $2R_{\rm PL}$ this would thus alone not make a difference for the effective flux.

Indeed, a more important correction comes from a proper treatment of the survival factor, which we do here exactly as in Ref.~\cite{Eskola:2022vpi} by calculating a Glauber-model probability for having no inelastic hadronic interaction between the nuclei in the optical approximation~\footnote{By using the total instead of inelastic nucleon-nucleon cross section, we are taking a probabilistic interpretation that elastic nucleon-nucleon interactions would also disintegrate the colliding nuclei.},
\begin{equation}
  \Gamma_{AB}^{\rm WS}({\bf b}) = \exp[-\sigma_{\rm NN} T_{AB}^{\rm WS}({\bf b})],
  \label{eq:glauber}
\end{equation}
where the nuclear overlap function
\begin{equation}
  T_{AB}^{\rm WS}({\bf b}) = \int {\rm d}^2{\bf s} \; T_{A}^{\rm WS}({\bf s}) T_{B}^{\rm WS}({\bf b} - {\bf s})
\end{equation}
is obtained from the WS distribution in Eq.~\eqref{eq:ws-distr} through
\begin{equation}
  T_{A}^{\rm WS}({\bf s}) = \int_{-\infty}^\infty {\rm d}z \; \rho_A^{\rm WS}\left(\sqrt{z^2+{\bf s}^2}\right)
  \label{eq:ws-ta}
\end{equation}
and the total hadronic nucleon-nucleon cross section is taken with a value $\sigma_{\rm NN} = 90\ {\rm mb}$ (cf.~Ref.~\cite{ParticleDataGroup:2016lqr}) at the considered $\sqrt{s_{\rm NN}} = 5.02\ {\rm TeV}$ collision energy. Here, the energy dependence of $\sigma_{\rm NN}$ takes into account an increase of the radius of nucleon-nucleon interactions at the LHC energies.

As an intermediate result, we keep the pointlike approximation for the target-nucleus PDF spatial dependence, $T_B^{\rm PL}({\bf s}) = B\delta^{(2)}({\bf s})$, in Eqs.~\eqref{eq:eff_flux}-\eqref{eq:eff_flux_b_dep} while employing the improvements described above. The effective flux obtained this way, labeled as WS$_{\delta({\bf s})}$, is shown with dashed lines in FIG.~\ref{fig:ph-flux} (b). We see that this gives a smoother transition around $2R_{\rm PL}$ compared to the PL approximation.

\subsection{Full Woods-Saxon effective flux}
\label{subsec:full_ws}

As a final step, we now account also for the spatial extent of the target-nucleus parton distributions by taking the nuclear thickness function in Eqs.~\eqref{eq:eff_flux}-\eqref{eq:eff_flux_b_dep} as the one calculated from the WS distribution, $T_B^{\rm WS}$, through Eq.~\eqref{eq:ws-ta}. The effective flux from this full WS calculation is presented in FIG.~\ref{fig:ph-flux} (b) with solid lines. Due to the additional smearing from the target-side spatial distribution, the effective flux now extends to much smaller values of $|{\bf r}|$ than in the other approximations. We see that this additional contribution is particularly important for high-energy photons for which the flux drops extremely fast as a function of $|{\bf r}|$.

Moreover, we can now clearly separate two regions: For $|{\bf r}| \agt 3R_{\rm PL}$ the impact parameter $|{\bf b}|$ is always larger than the sum of the two nuclear radii and thus the survival factor is one. In this region, the bare flux is also always well reproduced by the pointlike calculation. Thus, in this ``far-passing'' region (cf.\ FIG.~\ref{fig:transverse_plane}~(b)) the geometrical effects do not play a role and one has
\begin{equation}
  f_{\gamma/A}^{\rm eff}(y, |{\bf r}| \agt 3R_{\rm PL}) = f_{\gamma/A}^{\rm PL}(y, {\bf r})
\end{equation}
irrespective of the approximation used. On the contrary, for the $|{\bf r}| \alt 3R_{\rm PL}$ ``near-encounter'' events (cf.\ FIG.~\ref{fig:transverse_plane}~(c)) the three approximations differ from each other and the geometrical effects are important.

For completeness, we show in FIG.~\ref{fig:ph-flux}~(c) the effective flux also as a function of the impact parameter $|{\bf b}|$ between the nuclei. In the case of PL and WS$_{\delta({\bf s})}$ approximations this is of course identical to the flux as a function of $|{\bf r}|$, but for the full WS calculation we now see that the flux behaves differently as a function of $|{\bf r}|$ and $|{\bf b}|$. For the latter, one sees clearly how the survival factor $\Gamma_{AB}$ very effectively excludes any contribution from the region $|{\bf b}| < 2R_{\rm PL}$. Instead, the integration over finite ${\bf s}$ causes the effective flux at a fixed ${\bf b}$ to have a contribution from the region $|{\bf r}| < |{\bf b}|$ where the flux is larger than at $|{\bf r}| = |{\bf b}|$. Due to the non-linear behaviour, this increased contribution from $|{\bf r}| < |{\bf b}|$ is larger than the depletion from $|{\bf r}| > |{\bf b}|$ and thus the full WS flux as a function of $|{\bf b}|$ is always enhanced compared to the WS$_{\delta({\bf s})}$ approximation. For low-energy photons this enhancement is, however, small and WS$_{\delta({\bf s})}$ approximates well the full WS result at low photon energies. For the above reason, it is also not possible to separate the ``near-encounter'' and ``far-passing'' regions with a cut at a single value of $|{\bf b}|$.

\begin{figure}
  \centering
  \includegraphics{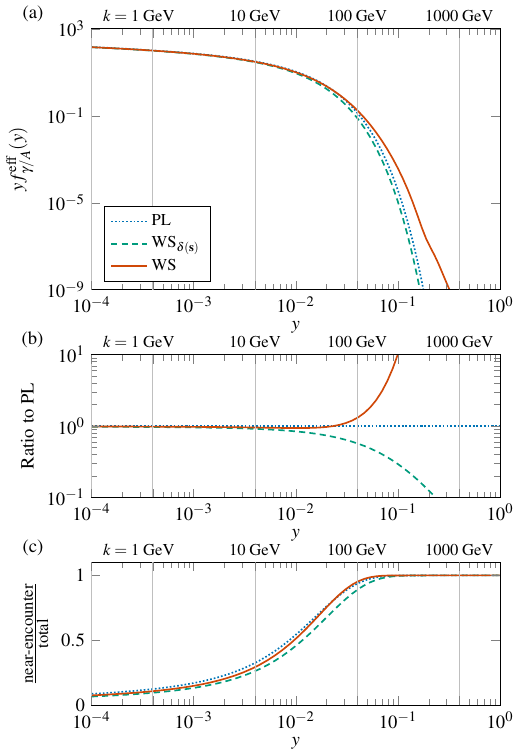}
  \caption{(a) The effective photon flux as a function of the momentum fraction $y$. Different values of the c.m.s.~frame photon energy $k = y\sqrt{s_{\rm NN}}/2$ are also indicated for $\sqrt{s_{\rm NN}} = 5.02\ {\rm TeV}$. (b) The ratio of WS and WS$_{\delta({\bf s})}$ approximations to the PL one. (c) The fraction of events coming from the ``near-encounter'' $|{\bf r}| < 3R_{\rm PL}$ region.}
  \label{fig:photon-flux-y-dep}
\end{figure}

Finally, FIG.~\ref{fig:photon-flux-y-dep}~(a) shows the effective flux as a function of the fraction $y$, integrated over all the transverse-space variables. For low-energy photons, $y \alt 10^{-3}$, the integrated flux comes almost entirely from the large-$|{\bf r}|$ ``far-passing'' events and thus the three approximations agree to a good precision, as seen in FIG.~\ref{fig:photon-flux-y-dep}~(b). As the photon energy grows, so does the fraction of flux from the small-$|{\bf r}|$ ``near-encounter'' configurations, which start to dominate at $y \sim 10^{-2}$, see FIG.~\ref{fig:photon-flux-y-dep}~(c), and the three approximations begin to divert. For $y \agt 10^{-1}$ the flux comes entirely from the ``near-encounter'' region but the flux also becomes quickly very small.

Note that we used here the same WS parametrization, Eq.~\eqref{eq:ws-distr}, for describing three different distributions: the nuclear charge distribution in calculating the EPA photon flux, the nucleon distribution within the nucleus in the hadronic-interaction survival factor $\Gamma_{AB}$, as well as the spatial distribution of partons in the target nucleus. Generally, these do not have to be the same, e.g.\ due to the neutron-skin effect and the finite size of the nucleons. We make no attempt here to take into account this subtlety.

\section{Inclusive UPC dijet photoproduction in Pb+Pb collisions at 5.02 TeV}
\label{sec:ATLAS-UPC-dijets}

For the small-$|{\bf r}|$ geometrical effects to be significant experimentally, we need an observable that is sensitive to them. This can be the case with the UPC dijet photoproduction, where the high $p_{\rm T}$ of the jets, in combination with a not-too-small rapidity of the dijet system in the photon-going direction, will require an energetic photon in the initial state and thus bias the collision towards smaller $|{\bf r}|$. We will now discuss this process in the context of the ATLAS measurement for Pb+Pb collisions at $\sqrt{s_{\rm NN}} = 5.02\ {\rm TeV}$~\cite{ATLAS:2022cbd}.

The ATLAS measurement uses the following fiducial cuts for the cross section definition: The jets used in defining the kinematical variables
\begin{equation}
  H_{\rm T} = \sum_{i \in {\rm jets}} p_{{\rm T}, i}, \quad z_\gamma = \frac{M_{\rm jets}}{\sqrt{s_{\rm NN}}} e^{+y_{\rm jets}}, \quad x_A = \frac{M_{\rm jets}}{\sqrt{s_{\rm NN}}} e^{-y_{\rm jets}}
\end{equation}
are reconstructed with the anti-$k_{\rm T}$ algorithm~\cite{Cacciari:2008gp} with $R = 0.4$ radius and are required to satisfy the condition
\begin{equation}
  p_{\rm T}^{\rm jet} > 15\ {\rm GeV}, \quad |\eta^{\rm jet}| < 4.4, \quad 35\ {\rm GeV} < M_{\rm jets} < 185\ {\rm GeV}.
  \label{eq:jet_cuts}
\end{equation}
Here, $p_{\rm T}^{\rm jet}$ and $\eta^{\rm jet}$ are the transverse momentum and pseudorapidity of a single jet and $M_{\rm jets}$ and $y_{\rm jets}$ are the invariant mass and rapidity of the $N \geq 2$-jet system passing the single-jet cuts, with the positive rapidity chosen to be in the photon-going direction. The photon direction is determined experimentally though additional event-level cuts. We will postpone the discussion of these additional event-selection criteria to Section~\ref{sec:bu-class} and the results in this section will be based on the cuts in Eq.~\eqref{eq:jet_cuts} only.

\begin{figure}
  \centering
  \includegraphics{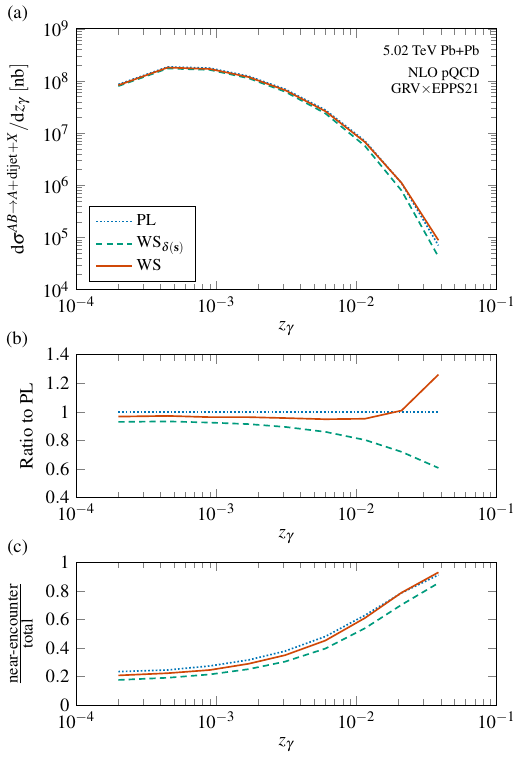}
  \caption{(a) The NLO pQCD cross section for the UPC dijet photoproduction in Pb+Pb collisions at $\sqrt{s_{\rm NN}} = 5.02\ {\rm TeV}$ within the ATLAS kinematics. (b) The ratio of WS and WS$_{\delta({\bf s})}$ approximations to the PL one. (c) The fraction of events coming from the ``near-encounter'' $|{\bf r}| < 3R_{\rm PL}$ region.}
  \label{fig:xsec-zg-dep}
\end{figure}

We calculate the cross section in NLO pQCD with the Frixione \& Ridolfi jet photoproduction code~\cite{Frixione:1997ks}. With the $M_{\rm jets} > 35\ {\rm GeV}$ cut, the leading-jet $p_{\rm T}$ is always larger than the minimum-$p_{\rm T}$ cut at the back-to-back two-jet limit, and therefore we are free from the large logarithmic corrections appearing with symmetric $p_{\rm T}$ cuts~\cite{Frixione:1997ks,Banfi:2003jj}. The fixed-order treatment as done here is thus valid for the Eq.~\eqref{eq:jet_cuts} cuts. We use the GRV para\-metrization~\cite{Gluck:1991jc} for the photon PDFs and take the Pb nPDFs from the EPPS21 analysis~\cite{Eskola:2021nhw} with CT18A NLO~\cite{Hou:2019efy} as the free-proton PDF baseline. We set the perturbative hard (renormalization and factorization) scale as $Q = \hat{H}_{\rm T}$, where
\begin{equation}
  \hat{H}_{\rm T} = \sum_{i \in {\rm partons}} p_{{\rm T}, i}
\end{equation}
is the parton-level version of the $H_{\rm T}$ variable with the sum going over all the (final-state) partons in the events accepted by the Eq.~\eqref{eq:jet_cuts} jet cuts.

FIG.~\ref{fig:xsec-zg-dep}~(a) shows the cross section as a function of $z_\gamma$. At leading order this variable would resolve to the product $x_\gamma y$ and hence the ${\rm d}\sigma^{AB \rightarrow A + {\rm dijet} + X} / {\rm d}z_\gamma$ distribution can be used to study the effective flux at different photon energies. Indeed, at small $z_\gamma$, dominated by low values of $y$, the three approximations discussed in the previous section give very similar predictions, with sub-10\% differences as shown in FIG.~\ref{fig:xsec-zg-dep}~(b). As $z_\gamma$ increases, so does the minimum value of $y$, and gradually a larger and larger contribution to the cross section comes from the ``near-encounter'' events, see FIG.~\ref{fig:xsec-zg-dep}~(c), where the three approximations differ significantly.

In the bin of highest $z_\gamma$, this leads to a 40\% suppression in the WS$_{\delta({\bf s})}$ prediction compared to the PL one. Even more dramatically, there is a factor of two difference between WS$_{\delta({\bf s})}$ and the full WS calculations, which shows that taking into account the full collision geometry, including the finite size of the target nucleus, is really needed for realistic predictions and correct interpretation of the measurement. The PL approximation works quite well in approximating the full WS calculation, except in this highest-$z_\gamma$ bin where the difference is of the order of 20\%~\footnote{Note that compared to our (preliminary) results presented in Ref.~\cite{Eskola:2023ioj}, we use here a narrower binning in $z_\gamma$, matching with that of Ref.~\cite{ATLAS:2022cbd}, which explains the slight differences in the quoted values.}\nocite{Eskola:2023ioj}. We note that the production of dijets in these highest values of $z_\gamma$ indeed requires the photon to have a c.m.s.~frame energy of at least 67.7 GeV, and the cross section in these kinematics is therefore truly probing the high-energy tail of the photon flux.

\section{Breakup-class modelling}
\label{sec:bu-class}

In addition to the jet cuts in Eq.~\eqref{eq:jet_cuts}, the ATLAS measurement uses the following event-selection criteria~\cite{ATLAS:2022cbd}
\begin{equation}
  0nXn, \quad \sum \Delta\eta_{0n} > 2.5, \quad \Delta\eta_{Xn} < 3.0.
  \label{eq:bu_cuts}
\end{equation}
The $0nXn$ condition, i.e.\ requiring that there are no forward neutrons observed in the zero-degree calorimeter (ZDC) in one direction and at least one ($X \geq 1$) neutron in the opposite direction, is needed in order to make sure that the photon-emitting nucleus stayed intact while the other broke up, suppressing the jet production from ($XnXn$) hadronic nucleus-nucleus and ($0n0n$) two-photon and diffractive processes, respectively. This allows also the photon-going direction to be identified as the one towards the $0n$-recording ZDC. In addition, $\sum \Delta\eta_{0n} > 2.5$ puts a lower limit on the sum of rapidity gaps in the photon-going direction and $\Delta\eta_{Xn} < 3.0$ requires that there is no large gap in the opposite direction, which are applied in order to further reduce the residual backgrounds from the above-mentioned processes.

While these cuts are necessary experimentally for a clean selection of photonuclear events, they pose a difficulty for the perturbative calculation. Our treatment in the previous section was essentially that for an $AnAn$ ($A \geq 0$) event class. Even though calculating the bare photon flux from the proper nuclear form factor ensured that the nucleus did not break up while emitting the photon that induced the hard interaction, we did not keep track of possible additional soft e.m.\ interactions that could excite and break the nucleus, and thus we implicitly allowed for any number of neutrons in the photon-going direction. Our treatment on the target side was also fully inclusive, thus $An$, implicitly including the diffractive contribution which would leave the target nucleus intact and yield zero neutrons in the target-going direction unless produced by the additional soft e.m.\ interactions. Furthermore, the rapidity-gap requirements reduce the allowed phase-space for soft radiation. Putting too strict cuts in the photon-going direction could also reduce the resolved contribution relative to the direct one. The use of a sum-of-gaps (instead of a one wide gap) condition should however alleviate these problems to some extent.

The assumption that we are making here is that the above cuts are \emph{inclusive enough} such that the pure NLO pQCD calculation as factorized in Eq.~\eqref{eq:xsec_full}, after taking into account the breakup-class modelling as discussed below, captures the main physics of this measurement. We also note that our predictions are for parton-level jets, i.e.\ we do not include any non-perturbative hadronization or underlying-event corrections which are to be studied with Monte-Carlo event generators~\cite{Helenius:2018mhx,Bierlich:2022pfr,Hoeche:2023gme}, where it also becomes possible to asses the potential sensitivity to the rapidity-gap cuts.

\begin{figure}
  \centering
  \includegraphics{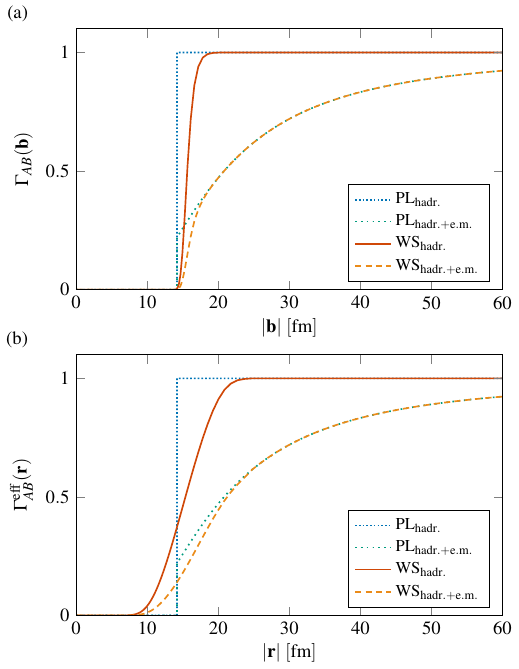}
  \caption{(a) The total survival probability without and with the e.m.\ breakup factor for the WS and PL approximations of the hadronic survival factor as a function of $|{\bf b}|$. (b) The effective survival factor as a function of $|{\bf r}|$.}
  \label{fig:survival-factor}
\end{figure}

To account for the $0n$ requirement in the photon-going direction, we follow the assumption that the soft-exitation probability factorizes from the hard interaction~\cite{Baltz:2002pp} and take the (Poissonian) probability for \emph{no} e.m.\ breakup of the nucleus $A$ from the Starlight generator~\cite{Klein:2016yzr} where it is given as
\begin{equation}
  \Gamma_{AB}^{\rm e.m.}({\bf b}) = \exp\left[ - \int_0^1 {\rm d}y f_{\gamma/B}(y,{\bf b}) \sigma_{\gamma A \rightarrow A^*}(\sqrt{y\,s_{\rm NN}}) \right]
  \label{eq:em_bu}
\end{equation}
with $\sigma_{\gamma A \rightarrow A^*}$ the photoexitation cross section, expressed here as a function of the photon-nucleon c.m.s.\ energy $\sqrt{s_{\rm \gamma N}} = \sqrt{y\,s_{\rm NN}}$. The total survival probability then takes the form
\begin{equation}
  \Gamma_{AB}^{\rm hadr.+e.m.}({\bf b}) = \Gamma_{AB}^{\rm e.m.}({\bf b}) \Gamma_{AB}^{\rm hadr.}({\bf b}),
\end{equation}
where $\Gamma_{AB}^{\rm hadr.}$ is the no-strong-interaction hadronic survival factor defined in Section~\ref{sec:eff-flux}.

FIG.~\ref{fig:survival-factor}~(a) shows a comparison of the survival factors obtained from the PL (i.e. hard-sphere) and WS approximations, without and with the e.m.\ survival factor. We see that the no e.m.\ breakup condition significantly lowers the survival probability, particularly at small impact parameters, but the full hadr.+e.m.\ factor still differs between the PL and WS approximations. In FIG.~\ref{fig:survival-factor}~(b), we study the effective $|{\bf r}|$-dependent survival factor (which multiplies the bare flux in Eq.~\eqref{eq:eff_flux_r_dep})
\begin{equation}
  \Gamma_{AB}^{\rm eff}({\bf r}) = \frac{1}{B} \int {\rm d}^2{\bf s} \, T_{B}({\bf s}) \, \Gamma_{AB}({\bf r}\!-\!{\bf s}),
\end{equation}
showing that there is still some phase-space left for the effective flux in the $|{\bf r}| < 2R_{\rm PL}$ region for the full WS approximation even after the suppression from the e.m.\ breakup is accounted for. We also see that, in absolute terms, the PL approximation gets a larger suppression than the WS one.

\begin{figure}
  \centering
  \includegraphics{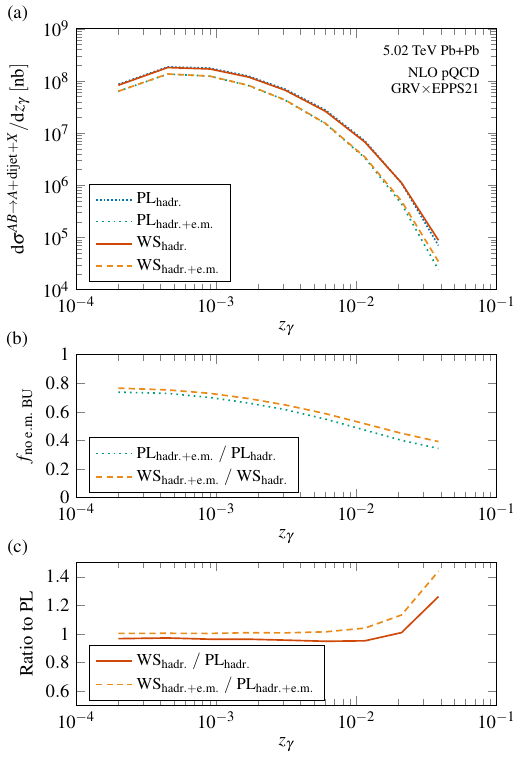}
  \caption{(a) The NLO pQCD cross section for the UPC dijet photoproduction in Pb+Pb collisions at $\sqrt{s_{\rm NN}} = 5.02\ {\rm TeV}$ within the ATLAS kinematics, without and with the e.m.\ breakup. (b) The e.m.\ breakup suppression factor from the PL and WS approximations of the effective flux. (c) The ratio of WS and PL predictions without and with the e.m.\ breakup.}
  \label{fig:xsec-zg-dep-w-bu-classes}
\end{figure}

The reduction in the survival probability yields also a visible impact in the dijet cross section, as shown in FIG.~\ref{fig:xsec-zg-dep-w-bu-classes}~(a). This effect can be quantified with the no e.m.\ breakup fraction $f_{\rm no\;e.m.\;BU}$, cf.\ Ref.~\cite{ATLAS:2022cbd}. Here, it is defined simply as the ratio between the NLO cross sections with and without the $\Gamma_{AB}^{\rm e.m.}$ included in calculating the effective flux. This fraction is shown in FIG.~\ref{fig:xsec-zg-dep-w-bu-classes}~(b), where we see that the effect ranges from a 20\% suppression at small $z_\gamma$ to 60\% at large $z_\gamma$, agreeing also with the experimental estimate in Ref.~\cite{ATLAS:2022cbd}, particularly at large $z_\gamma$. The effect is thus quite significant and has to be accounted for in the data-to-theory comparisons. The PL and WS cross sections, where the latter refers to the full one with all the geometrical effects included, receive about the same suppression, the PL having only slightly larger one from a stronger reduction in the survival factor. This in fact leads to an enhancement in the WS to PL ratio, as shown in FIG.~\ref{fig:xsec-zg-dep-w-bu-classes}~(c), where the WS prediction now has an about 40\% enhancement compared to the PL one in the bin of highest $z_\gamma$ after taking into account the e.m.\ suppression factor.

We note that even after accounting for the e.m.\ breakup fraction as in Eq.~\eqref{eq:em_bu}, our calculation does not agree exactly with the $0nXn$ condition. To achieve it, we should subtract the diffractive photonuclear contribution which is still implicitly included in our inclusive calculation \footnote{For example, PDFs measured in fully inclusive deep-inelastic scattering do include also a diffractive contribution.}. A part of this contribution could actually survive the $0nXn$ event selection due to the target nucleus breaking up from the additional soft e.m.\ interactions, but these events are likely excluded by the no-gap requirement. In a very strict sense, this means that the measurement with the cuts in Eq.~\eqref{eq:bu_cuts} is not directly probing the inclusive nPDFs, but rather a certain difference of the inclusive and diffractive nPDFs. The diffractive contribution is, however, expected to be significant only at small $x_A$ (i.e.\ large $z_\gamma$), where it has been estimated to be at a sub-10\% level within the ATLAS measured range $x_A > 10^{-3}$, quickly diminishing towards higher values of $x_A$~\cite{Guzey:2020ehb}. We thus believe, as stated above, that our WS$_{\rm hadr.+e.m.}$ prediction should match with the experimental cross section with the fiducial cuts given in Eqs.~\eqref{eq:jet_cuts} and~\eqref{eq:bu_cuts} to a good precision, and we do not perform this (currently model dependent and experimentally unconstrained) subtraction here. Further studies in the $XnXn$ and $0n0n$ event classes would certainly aid in putting the modelling of the e.m.\ survival factor and the diffractive contribution on a firmer ground~\cite{ATLAS:2022cbd}.

\section{Spatial dependence of nuclear modifications}
\label{sec:spatial-npdfs}

Thus far, we have used the factorization assumption of Eq.~\eqref{eq:b_dep_fact_assump} in our calculations. In this approach the spatial degrees of freedom are taken to be fully decoupled from the longitudinal-momentum-fraction dependence of the nPDFs. Thus each nucleon within the nucleus is assumed to have the same nuclear modifications irrespectively of the transverse position. This is, however, against the physical intuition of e.g.\ the nuclear shadowing where the local nuclear thickness is certainly important~\cite{Armesto:2006ph,Frankfurt:2011cs}. In this section we relax the previous assumption and study the corrections arising from the spatial dependence of the nuclear modifications by applying the phenomenological model described in Ref.~\cite{Helenius:2012wd}.

We begin by re-writing Eq.~\eqref{eq:xsec_full}, in a fully general form, as
\begin{multline}
  {\rm d} \sigma^{AB \rightarrow A + {\rm dijet} + X} \\
  = \sum_{i,j,X'} \int {\rm d}^2{\bf s} \, f_{\gamma/A}^{\rm eff}(y,{\bf s}) \otimes f_{i/\gamma}(x_\gamma,Q^2) \otimes f_{j/B}(x,Q^2,{\bf s}) \\
  \otimes {\rm d} \hat{\sigma}^{ij \rightarrow {\rm dijet} + X'}(x_\gamma y p_A,x p_B,Q^2)
  \label{eq:xsec_re}
\end{multline}
where
\begin{equation}
  f_{\gamma/A}^{\rm eff}(y,{\bf s}) = \int {\rm d}^2{\bf r} \, \Gamma_{AB}({\bf r}\!-\!{\bf s}) \, f_{\gamma/A}(y,{\bf r})
  \label{eq:eff_flux_s_dep}
\end{equation}
is the ${\bf s}$-dependent effective flux. This function now sets how the partons in nucleus $B$ are sampled in the transverse space. If it would be constant over the support of $f_{j/B}(x,Q^2,{\bf s})$, the ${\bf s}$-integral in Eq.~\eqref{eq:xsec_re} could be taken trivially and one would recover the spatially averaged nPDFs. Conversely, if $f_{\gamma/A}^{\rm eff}(y,{\bf s})$ is \emph{not} a constant within this region, this would indicate that the nucleus $B$ partons at different transverse locations can feel different effective flux, whereby the observable would be sensitive to the spatially-dependent nPDFs. Note that one would recover the effective flux in Eq.~\eqref{eq:eff_flux} with
\begin{equation}
  f_{\gamma/A}^{\rm eff}(y) = \frac{1}{B} \int {\rm d}^2{\bf s} \, f_{\gamma/A}^{\rm eff}(y,{\bf s}) \, T_{B}({\bf s}).
\end{equation}

\begin{figure}
  \centering
  \includegraphics{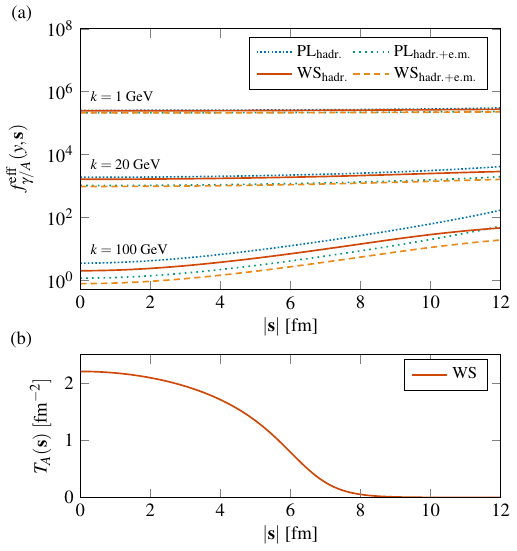}
  \caption{(a) The effective flux of Eq.~\eqref{eq:eff_flux_s_dep} as a function of the transverse distance from the center of the target in Pb+Pb collisions at $\sqrt{s_{\rm NN}} = 5.02\ {\rm TeV}$. (b) The nuclear thickness function for the Pb target.}
  \label{fig:ph-flux-s-dep}
\end{figure}

The ${\bf s}$-dependent effective flux is shown in FIG.~\ref{fig:ph-flux-s-dep}~(a) for the PL and WS approximations, without and with the e.m.\ breakup modelling. As could be expected from the discussion in the previous sections, for soft photons the ${\bf s}$ dependence is almost flat and thus cannot probe the spatial distribution of the target. For very energetic photons with $k = 100\ {\rm GeV}$, we however see non-negligible ${\bf s}$ dependence, and from the rising trend as a function of $|{\bf s}|$, we find that these photons sample the nuclear partons more from the edge of the target nucleus than from its center, as can be verified by comparing with the nuclear thickness function shown in FIG.~\ref{fig:ph-flux-s-dep}~(b). This enhancement towards the edge is yet another way to understand the enhancement in the full WS prediction compared to the WS$_{\delta({\bf s})}$.

To quantify the sensitivity to the spatially dependent nPDFs, we now generalize Eq.~\eqref{eq:b_dep_fact_assump} to
\begin{equation}
  f_{j/B}(x,Q^2,{\bf s}) = \frac{1}{B} \, T_{B}({\bf s}) \sum_N r_{j/N/B}(x,Q^2,{\bf s}) f_{j/N}(x,Q^2),
\end{equation}
where the sum goes over the nucleons $N \in \{p,n\}$ in nucleus $B$, with $f_{j/N}$ being the free-nucleon PDFs, and $r_{j/N/B}$ are the spatially dependent nuclear modifications. In EPS09s~\cite{Helenius:2012wd}, these modifications are parametrized in terms of a power series of the nuclear thickness function, terminated at power four,
\begin{equation}
  r_{j/N/B}(x,Q^2,{\bf s}) = \sum_{m=0}^4 c_m^{j/N}(x,Q^2) [T_{B}({\bf s})]^m,
\end{equation}
with the condition $c_0^{j/N}(x,Q^2) \equiv 1$ ensuring that the nuclear modifications vanish at the large-$|{\bf s}|$ limit.

With this parametrization, the cross section factorizes as
\begin{multline}
  {\rm d} \sigma^{AB \rightarrow A + {\rm dijet} + X} \\= \sum_{i,j,X'} \sum_{m=0}^4 f_{\gamma/A}^{{\rm eff},m}(y) \otimes f_{i/\gamma}(x_\gamma,Q^2) \otimes f_{j/B}^m(x,Q^2) \\
  \otimes {\rm d} \hat{\sigma}^{ij \rightarrow {\rm dijet} + X'}(x_\gamma y p_A,x p_B,Q^2)
  \label{eq:xsec_EPS09s}
\end{multline}
with the generalised ($m$-dependent) effective flux
\begin{equation}
  \begin{split}
    f_{\gamma/A}^{{\rm eff},m}(y) &= \frac{1}{B} \int {\rm d}^2{\bf r} \int {\rm d}^2{\bf s} \, f_{\gamma/A}(y,{\bf r}) \, [T_{B}({\bf s})]^m \, \Gamma_{AB}({\bf r}\!-\!{\bf s}) \\
    &= \frac{1}{B} \int {\rm d}^2{\bf s} \, f_{\gamma/A}^{\rm eff}(y,{\bf s}) \, [T_{B}({\bf s})]^m,
    \label{eq:eff_flux_m_dep}
  \end{split}
\end{equation}
and the nPDFs with EPS09s coefficients
\begin{equation}
  f_{j/B}^m(x,Q^2) = \sum_N c_m^{j/N}(x,Q^2) f_{j/N}(x,Q^2)
\end{equation}
again being easily implementable in the jet photoproduction codes.

\begin{figure}
  \centering
  \includegraphics{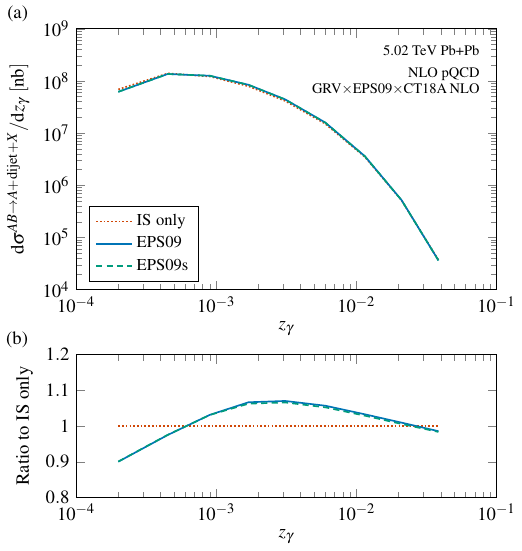}
  \caption{(a) The NLO pQCD cross section for the UPC dijet photoproduction in Pb+Pb collisions at $\sqrt{s_{\rm NN}} = 5.02\ {\rm TeV}$ within the ATLAS kinematics with results using no nuclear modifications in the nucleon PDFs (IS only), with the EPS09 modifications, or using the EPS09s spatially dependent modifications. (b) Ratio to the prediction without nuclear modifications.}
  \label{fig:xsec-zg-dep-w-EPS09s}
\end{figure}

FIG.~\ref{fig:xsec-zg-dep-w-EPS09s}~(a) shows the UPC dijet cross section within the full WS$_{\rm hadr.+e.m.}$ approximation of the flux as defined in the previous section. The GRV photon PDFs are used as before and the free-nucleon PDFs $f_{j/N}$ are taken from the CT18A NLO analysis assuming the isospin symmetry between the neutrons and protons as with the EPPS21 nPDFs in the previous sections. Three results are compared here, with the isospin (IS) only referring to the case with no nuclear modifications, i.e.\ $r_{j/N/B}(x,Q^2,{\bf s}) \equiv 1$. In the second option, the spatially independent EPS09 nuclear modifications~\cite{Eskola:2009uj} are applied, which matches the factorization assumption of Eq.~\eqref{eq:b_dep_fact_assump} as used in the previous sections. Finally, the spatially dependent modifications with the EPS09s $c_m^{j/N}$ coefficients are applied using the factorization in Eq.~\eqref{eq:xsec_EPS09s}.

\begin{figure}
  \centering
  \includegraphics{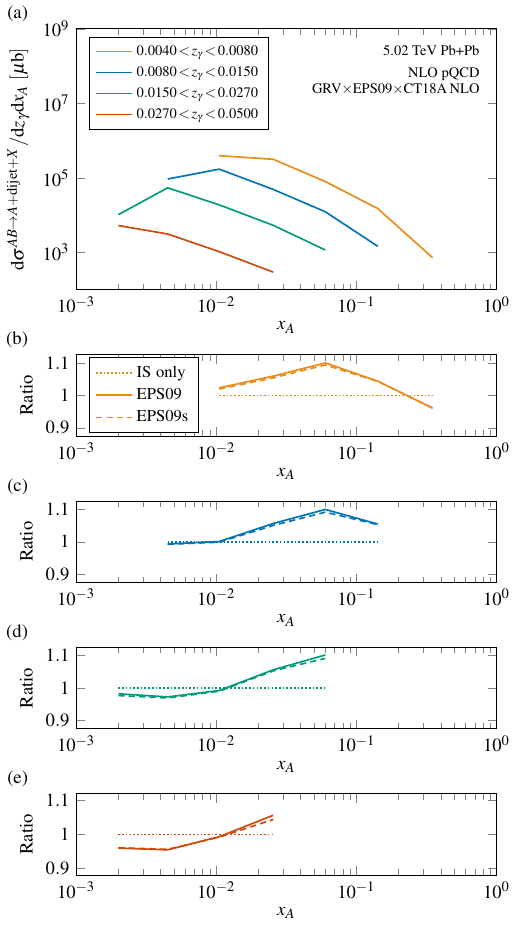}
  \caption{(a) The double-differential NLO pQCD cross section for the UPC dijet photoproduction in Pb+Pb collisions at $\sqrt{s_{\rm NN}} = 5.02\ {\rm TeV}$ within the ATLAS kinematics with results using no nuclear modifications in the nucleon PDFs (IS only), with the EPS09 modifications, or using the EPS09s spatially dependent modifications. (b-e) Ratios to the prediction without nuclear modifications.}
  \label{fig:xsec-xA-dep-w-EPS09s}
\end{figure}

Comparing the results by taking the ratio to the IS-only prediction in FIG.~\ref{fig:xsec-zg-dep-w-EPS09s}~(b), we see that the nuclear modifications from EPS09 are in the 10\% range (cf.\ Refs.~\cite{Guzey:2018dlm,Helenius:2018mhx}). Based on the results in FIG.~\ref{fig:ph-flux-s-dep}, we should expect the EPS09s to match with EPS09 at small $z_\gamma$ and then, due to the larger flux at the edge where the nuclear modifications are smaller, the EPS09s curve should be between the EPS09 and IS-only predictions for larger $z_\gamma$. This is indeed seen in FIG.~\ref{fig:xsec-zg-dep-w-EPS09s}~(b), but the correction from the spatial dependence of the nuclear modifications is extremely small. At the highest values of $z_\gamma$ this could be a result of corrections from shadowing and antishadowing regions cancelling each other, so we plot in FIG.~\ref{fig:xsec-xA-dep-w-EPS09s}~(a) the double-differential dijet cross section as a function of $x_A$ in the four largest-$z_\gamma$ bins. By looking again at the ratios to IS only, presented in FIG.~\ref{fig:xsec-xA-dep-w-EPS09s}~(b-e), we see that the difference between EPS09 and EPS09s is small for all values of $x_A$. This confirms that while the spatial dependence of the effective flux at high photon energies was able to resolve the overall shape of the target nucleus in the UPC dijet production, this spatial resolution is not strong enough to probe the spatial dependence of the nuclear modifications. Pushing the measurement to even larger values of $z_\gamma$ could in principle resolve more details, but due to the rapidly vanishing photon flux, it is certainly very difficult experimentally to gather sufficient statistics for this.

\section{Summary and discussion}
\label{sec:summary}

We have presented here a detailed assessment of the geometrical effects in hard inclusive UPC dijet photoproduction. Our NLO pQCD calculation takes for the first time into account the finite-size effects from both the photon-emitting and the target nucleus. By comparing these results to the well-known pointlike approximation, we have shown that the UPC dijet photoproduction cross section is sensitive to the transverse-plane geometrical effects which lead to a 20\% enhancement in the dijet cross section at large $z_\gamma$. After taking into account the forward-neutron-class modelling as described in Section~\ref{sec:bu-class}, this correction grew to 40\% in this highest-$z_\gamma$ bin. We also found that the leading finite-size effects obtained through factorizing the transverse and longitudinal degrees of freedom as in Eq.~\eqref{eq:b_dep_fact_assump} were sufficient for an accurate prediction and the spatial dependence of nuclear modifications as parametrized in EPS09s yielded only a small correction that is beyond the expected experimental precision.

This spatial sensitivity was a result of meeting the following conditions:
\begin{enumerate}
  \item The requirement of a UPC, which restricts the impact-parameter space by excluding the region with nuclear overlap.
  \item The non-linear and steeply falling behaviour of the flux for high-energy photons as a function of the distance from the emitter.
  \item The large enough size of the target nucleus across which the flux can vary.
  \item The requirement of a high-mass final state, which sets a (rapidity-dependent) lower limit for the allowed initial-state photon energy.
\end{enumerate}
These conditions are not specific to the dijet photoproduction but can be fulfilled also in production of other final states. This would, in principle, open a door to study the impact-parameter dependent nPDFs with a more general set of inclusive UPC observables. We note that in Ref.~\cite{Adeluyi:2011rt} it was also quoted that the difference between the analytic pointlike effective flux and a one averaged over the target nucleus transverse area was of the order of 10--15\% for the direct photoproduction of heavy quarks and already in Ref.~\cite{Baron:1993nk} these geometrical effects were shown to be significant for charm-quark pair production at high enough invariant mass of the pair.

As stated in Section~\ref{sec:impact-par-dep-factorization}, the impact-parameter dependent PDFs can also be understood as the Fourier transforms of the (off-forward) \emph{skewless} GPDs~\cite{Burkardt:2000za,*Burkardt:2002hr}. There is thus a very interesting connection between the hard inclusive UPCs and exclusive processes where these GPDs are usually studied. One might note, however, that this connection arrises from the fact that the inclusive UPC cross sections are not \emph{fully} inclusive. Without the UPC condition (i.e.\ with $\Gamma_{AB} \equiv 1$), all the spatial dependence would vanish straight from Eq.~\eqref{eq:xsec_full}. It is thus extremely important for an accurate interpretation of the observable to have a matching description of this condition on the experimental and theory sides, which includes also the proper treatment of the forward-neutron-class requirement. This breakup-class selection required additional modelling for the theory calculations and, as we discussed in Section~\ref{sec:bu-class}, poses a difficulty for the inclusive interpretation of the data, paralleling with the statement above. The measurements in the $0n0n$ and $XnXn$ event classes, as planned~\cite{ATLAS:2022cbd}, will be very helpful in quantifying the magnitude of the diffractive contribution and the Coulomb-exitation breakup fraction.

In addition to the nuclear effects discussed here, the photoproduction process depends on the partonic structure of resolved photons which is currently not well constrained and provides an another source of theoretical uncertainty. Throughout the article we have assumed the colliding nuclei to be spherically symmetric. For deformed nuclei the calculation would become even more involved as then one would need to keep track of the relative orientation of the nuclei. Furthermore, our mean-field optical-Glauber treatment of the hadronic-interaction probability neglects the initial-state fluctuations and nucleon-nucleon correlations in the nuclear wave function~\cite{Baym:1995cz}, which determine the event-by-event nucleon transverse-plane positions. To include these initial-state fluctuations, one should for consistency also take into account the changes in the photon flux due to the fluctuating nucleon positions~\cite{Bzdak:2011yy} and the effect of event-by-event fluctuating spatially dependent nPDFs, such as proposed recently in Ref.~\cite{Kuha:2024kmq}. We leave such considerations for future work.

\begin{acknowledgments}
  We thank Nestor Armesto for his help regarding the Frixione \& Ridolfi jet photoproduction code and Benjamin Gilbert and Blair Seidlitz for discussions concerning the ATLAS measurement. This research was funded through the Research Council of Finland projects No.\ 330448 and No.\ 331545, as a part of the Center of Excellence in Quark Matter of the Research Council of Finland (projects No.\ 346325, No.\ 346326, No.\ 364192 and No.\ 364194) and as a part of the European Research Council project ERC-2018-ADG-835105 YoctoLHC. We acknowledge computing resources from the Finnish IT Center for Science (CSC), utilised under the project jyy2580.
\end{acknowledgments}

\appendix

\section*{Appendix: Technical implementation}

As discussed in the main text, by factoring the transverse-plane integrals into the effective flux [Eqs.~\eqref{eq:eff_flux} and~\eqref{eq:eff_flux_m_dep}], it is relatively easy to use the traditional jet photoproduction codes for calculating the cross sections also in the impact-parameter-dependent factorization. The numerical evaluation of the full-WS flux in these calculations is somewhat involved, and we give some technical details of our implementation in this appendix. As noted in Ref.~\cite{Eskola:2022vpi}, the integrand in Eq.~\eqref{eq:ws-bare-flux} oscillates rapidly at large $|{\bf r}|$ and the integral is therefore poorly converging. For this, we use the result that at large distances the bare flux always reduces to that of a PL source and substitute for the large-$|{\bf r}|$ bare flux
\begin{equation}
  f_{\gamma/A}^{\rm WS}(y, |{\bf r}| > 2R_{\rm PL}) = f_{\gamma/A}^{\rm PL}(y, {\bf r}).
\end{equation}
Furthermore, for smaller $|{\bf r}|$ we reduce the number of numerical integrations by using the analytic expression for the Fourier transform of the WS distribution given in Ref.~\cite{Maximon:1966sqn}.

\begin{figure}
  \centering
  \includegraphics{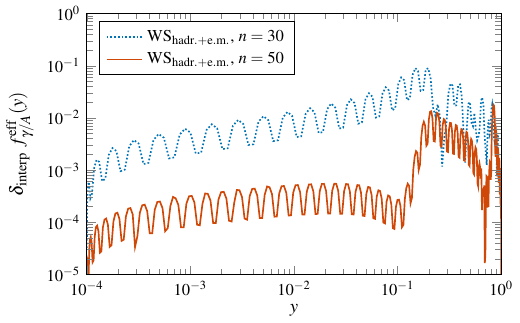}
  \caption{Performance of the chosen interpolation routine for different numbers of interpolation points.}
  \label{fig:inter-error}
\end{figure}

Solving the effective flux still requires calculating nested transverse-space integrals numerically. Doing this separately for every event in the Monte-Carlo implementation of the jet photoproduction calculation becomes quickly infeasible and one is forced to use some interpolation routine instead. Inspired by Ref.~\cite{Diehl:2021gvs}, we use the barycentric Lagrange interpolation with Chebyshev interpolation points~\cite{Berrut:2004}. In practice, we find best results by doing this in a $\sqrt[4]{y}$--$\log f_{\gamma/A}^{\rm eff}(y)$ space in order to have sufficient precision both at small and large values of $y$. The interpolating function then becomes
\begin{equation}
  f^{\rm interp}_n(y) = \exp\left[ \left( \sum_{j=0}^n \frac{w_j \log f_j}{\sqrt[4]{y_j} - \sqrt[4]{y}} \right) \bigg/ \left( \sum_{j=0}^n \frac{w_j}{\sqrt[4]{y_j} - \sqrt[4]{y}} \right) \right], \label{eq:barycentric}
\end{equation}
where the log-flux values
\begin{equation}
  \log f_j = \log f_{\gamma/A}^{\rm eff} (y_j)
\end{equation}
are sampled at the Chebyshev points of the second kind on the interval $y_j \in [10^{-4}, 1]$ in the $\sqrt[4]{y}$ space,
\begin{equation}
  \sqrt[4]{y_j} = 0.55 - 0.45 \cos\frac{j\pi}{n}, \qquad j = 0, \ldots, n,
\end{equation}
and the corresponding weights are~\cite{Berrut:2004}
\begin{equation}
  w_j = (-1)^j \delta_j, \qquad \delta_j = \begin{cases}
    1/2, & j = 0 \text{ or } j = n, \\
    1, & \text{otherwise}.
  \end{cases}
\end{equation}

As explained in Ref.~\cite{Berrut:2004}, Eq.~\eqref{eq:barycentric} can be evaluated in $\mathcal{O}(n)$ time. We find that with $n = 50$, the interpolation error
\begin{equation}
  \delta_{\rm interp}\ f_{\gamma/A}^{\rm eff}(y) = \left| 1 - f^{\rm interp}_n(y) / f_{\gamma/A}^{\rm eff}(y) \right|
\end{equation}
stays for the worst-performing case (WS$_{\rm hadr.+e.m.}$) below one per mille at $y < 10^{-1}$ and is about one percent at maximum in the region above that, see FIG.~\ref{fig:inter-error}, which is sufficient for our needs.

\bibliography{paper_PRC_resub}

\end{document}